\documentclass[12pt]{article}
\usepackage[utf8]{inputenc}
\usepackage[T1]{fontenc}
\usepackage{amsmath}
\usepackage{amsfonts}
\usepackage{amssymb}
\usepackage{graphicx}
\usepackage{float}
\usepackage{caption}
\usepackage{subcaption}
\usepackage{geometry}
\usepackage{cite}
\usepackage{url}
\usepackage{hyperref}
\usepackage{booktabs}
\usepackage{multirow}
\usepackage{array}
\usepackage{braket} %
\usepackage[utf8]{inputenc}
\usepackage[T1]{fontenc}
\usepackage{amsmath}
\usepackage{amssymb}
\usepackage[ruled,vlined,linesnumbered]{algorithm2e}
\usepackage{hyperref}
\usepackage{physics}  %
\usepackage[T1]{fontenc}
\usepackage{amsmath}
\usepackage{amssymb}
\usepackage{authblk}
\usepackage{enumitem}
\title{Quantum reinforcement learning-based active flow control}

\author[1]{Hongfu Zhang}
\author[1]{Hui Tang\thanks{Corresponding author: h.tang@polyu.edu.hk}}

\affil[1]{Department of Mechanical Engineering, The Hong Kong Polytechnic University, Hong Kong}

\begin{document}

\maketitle

\begin{abstract}
Active flow control remains a significant challenge due to the high-dimensional, nonlinear nature of fluid dynamics. Quantum machine learning may prove effective in addressing these issues, given that quantum computing possesses superiority over traditional computing in some extend. Thus, this study developed a quantum reinforcement learning (QRL) based active flow control framework, integrating variational quantum circuits (VQCs) with the proximal policy optimization (PPO) algorithm to learn a real time controller. Firstly, we tested the QRL in a CartPole problem. The QRL shows parameter efficiency and enhanced learning capability, indicating VQC acts as promising candidates for advancing RL, particularly in scenarios requiring both computational efficiency and robust performance. The active control of flow past a square circular cylinder at a Reynolds number of 100 was tested via QRL. Our hybrid architecture encodes high-dimensional flow states into a quantum policy network, which generates continuous blowing/suction actions on the cylinder’s surface, and thus suppress the vortex shedding to achieve drag reduction. Numerical simulations demonstrate the QRL successfully reduces the mean drag and attenuates lift oscillations. Flow field analysis confirms that QRL control effectively suppresses large-scale vortex shedding, leading to a narrower, more stable wake compared to the uncontrolled baseline. These results validate the potential of quantum-enhanced learning for tackling complex fluid dynamics problems. The proposed QRL framework establishes a promising blueprint for quantum-AI accelerated solutions in aerospace design, energy-efficient turbomachinery, and other applications involving sophisticated fluid-structure interactions.
\end{abstract}

\section{Introduction}
Fluid dynamics faces a persistent challenge in mitigating unsteady drag forces generated by flow separation around bluff bodies, particularly circular cylinders. Under typical subcritical Reynolds numbers, such flows exhibit periodic vortex shedding, known as the Kármán vortex street, inducing high oscillatory lift and drag forces, structural vibrations, and acoustic noise. Therefore, actively manipulating flow characteristics through control strategies holds significant appeal across various fluid mechanics applications, with the potential to yield substantial industrial benefits. However, deriving real-time, optimal control policies remains exceptionally difficult due to the chaotic, high-dimensional nature of the Navier-Stokes equations, sparse delayed rewards (where control actions impact flow states over extended timescales), and the need to adapt to transient features like separation points and shear-layer instabilities. Active Flow Control (AFC) has undergone rapid expansion and now plays a pivotal role in advancing industrial and sustainable solutions. AFC, via localized surface blowing/suction and control algorithm, offers substantial promises for suppressing vortex formation and reducing mean drag. For example, Flinois et al. (2015) established an optimal control framework based on adjoint methods to effectively stabilize vortex shedding. Leclercq et al. (2019) introduced a feedback-loop strategy employing iteratively linearized models to mitigate oscillations in resonating flows. For controlling the flow around a circular cylinder, Bergmann et al. (2005) derived an optimal control method utilizing reduced-order models built on proper orthogonal decomposition. Brackston et al. (2016) designed a feedback controller via a stochastic modelling approach and demonstrated its efficacy experimentally, successfully suppressing the asymmetric large-scale structure in the wake of a bluff body equipped with active flaps. More advanced methods, such as Proportional-Integral-Derivative controllers (Razvarz et al., 2019) and model-predictive control (Ghiglieri and Ulbrich, 2014) were also developed. Typically, such methods rely heavily on simplified models, which are called model-based control. They rely on harmonic or constant forcing mechanisms (Brunton and Noack, 2015; Schoppa and Hussain, 1998) or lookup tables that fail to generalize across unsteady flow regimes.

In contrast, model-free control strategies, which derive control policies directly from data using learning-based methodologies, show significant promise for application in complex, high-dimensional, nonlinear systems (Brunton and Noack, 2015; Garnier et al., 2021). The principal techniques within this paradigm include genetic algorithms (GAs) and artificial neural networks (ANNs). Although GAs have a well-established history of application in AFC (Chen and Aubry, 2005; Li et al., 2006; Wang et al., 2011; Zhang et al., 2013), ANNs are receiving growing interest, driven by accelerated progress in artificial intelligence and machine learning. Furthermore, previous studies indicate that ANNs can handle tasks of greater complexity and achieve higher learning speeds compared to GAs (Krizhevsky et al., 2017; LeCun et al., 2015). Among other machine learning techniques, the integration of ANNs with reinforcement learning algorithms has generated substantial attention (Mnih et al., 2015; Sutton and Barto, 2018). This fusion, known as deep reinforcement learning (DRL), has successfully addressed several challenging problems, such as mastering diverse Atari games without pre-programmed strategies (Mnih et al., 2015), generating realistic dialogues (Vinyals and Le, 2015), and managing the dynamics of sophisticated robots (Levine et al., 2018). Compared to data-driven supervised learning approaches, also applied in fluid mechanics for tasks like particle image velocimetry (PIV) measurement (Schanz et al., 2016; Wieneke, 2017), reduced-order modeling (Tai et al., 2021), and flow feature prediction (Brunton et al., 2020; Duraisamy et al., 2019), DRL possesses the distinct advantage of discovering solutions through trial-and-error, without requiring prior knowledge of the solution. It is particularly noteworthy that complex systems successfully controlled by DRL often exhibit nonlinearity and high-dimensionality, characteristics that are strikingly similar to the challenging nature of flow phenomena in AFC. Consequently, DRL is considered a highly promising approach for performing AFC.

However, these methods suffer from poor generalization performance and may convergence to suboptimal policies due to the curse of dimensionality and sparse reward landscapes (Viquerat et al., 2022). The limitations of classical DRL underscore the need for paradigm-shifting computational approaches. Hybrid architecture, combining parameterized variational quantum circuits (VQCs) with classical deep learning, introduces transformative capabilities ideally suited for flow control. VQC is composed of Quantum Neural Networks (QNNs). By leveraging quantum parallelism, a n-qubit QNN can simultaneously process $2^n$ high-dimensional states (e.g., pressure/velocity fields mapped via quantum feature embedding), exponentially accelerating state representation compared to classical neural networks. Furthermore, QNNs has theoretically and experimentally demonstrated a notable potential to build models with strong generalization capabilities even when trained on relatively small datasets (Caro et al., 2022).  Crucially, QNNs naturally model stochastic dynamics through superposition and entanglement, which could provide inherent robustness against chaotic noise in turbulent flows—a critical advantage where classical networks overfit. Recent Quantum Reinforcement Learning (QRL) frameworks demonstrate that they could perform better on a large search space, learn faster and better balance between exploration and exploitation compared with classical DRL (Caro et al., 2022; Li et al., 2020). Theoretically, QRL was shown to achieve quadratic improvements in learning efficiency and exponential improvements in performance for a broad class of learning problems (Dong et al., 2008; Li et al., 2020). This algorithm was also generalized to better adjust weights on favorable actions, which further demonstrated the robustness of this framework (Dong et al., 2008; Dong et al., 2012; Dunjko et al., 2016; Li et al., 2020). This capability aligns well with AFC scenarios where optimal actions, such as a time-dependent actuator to disrupt vortex structures, rely on historical flow states.
This paper pioneers the integration of VQCs with RL for real-time active control for drag reduction of a square clyinder, developing a QRL-based flow control framework. Our algorithm encodes CFD state snapshots (e.g., near-wall pressure distributions, velocity fields) into quantum states using a variational encoding circuit, while a parameterized quantum policy network, constructed from rotational gates and entanglement layers, generates continuous blowing/suction actions. The method will establish a scalable blueprint for quantum-AI acceleration in complex fluid-structure interaction problems, spanning aerospace design, energy-efficient turbomachinery, and next-generation smart structures.

\section{Methodology}

\subsection{AFC-oriented quantum neural network}
VQC, also referred to as parameterized quantum circuits, constitute a specialized category of quantum circuits that incorporate trainable parameters. These circuits incorporate tunable parameters, making them adaptable and trainable. The development of purely VQC-based QRL faces significant challenges due to its high demand for qubit resources. In AFC problems, the dimensionality of the complex flow dynamic state is typically significantly high which needs lots of qubits. To overcome this limitation, the present study aims to design a quantum neural network model that integrates a VQC with a classical ANN, where the ANN is employed to perform dimensionality reduction.

\begin{figure}[H]
\centering
\includegraphics[width=1.0\textwidth]{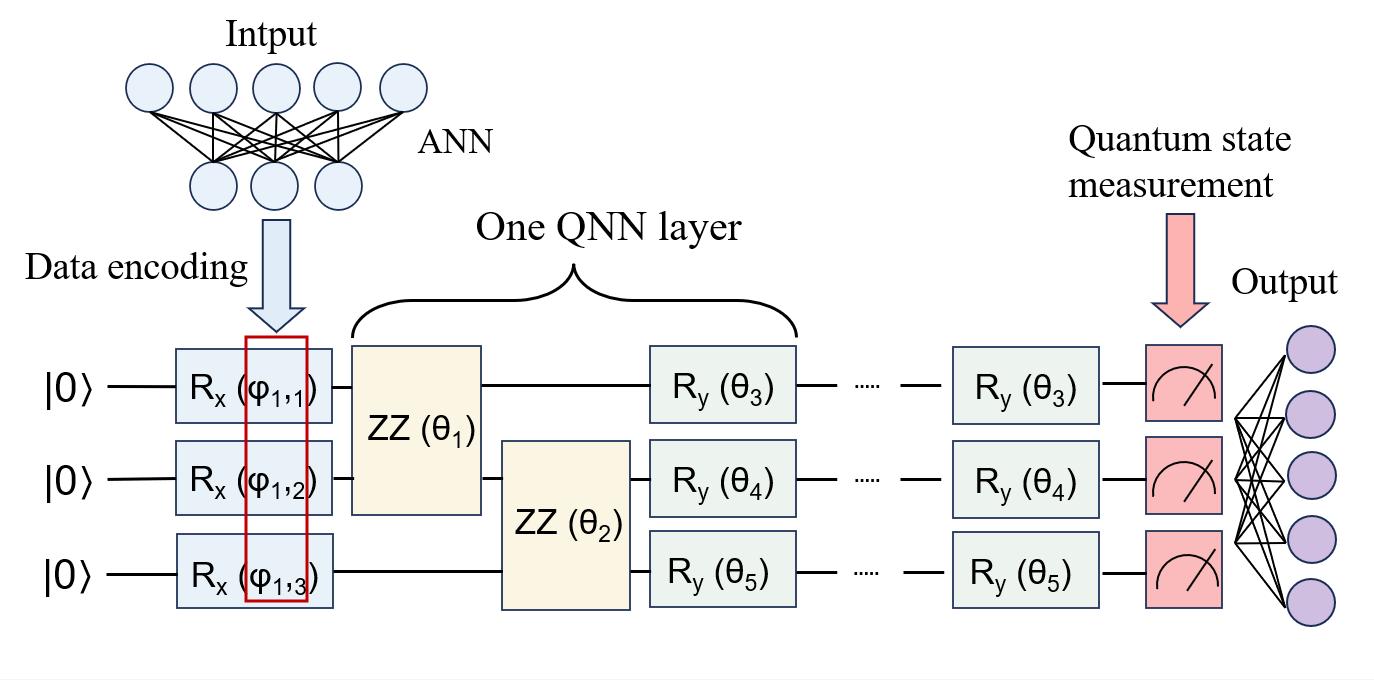}
\caption{Hybrid quantum-classical framework}
\label{fig:a1}
\end{figure}

The operation of quantum neural network model can be divided into 3 steps. The first one is quantum state preparation step. The input classical data is encoded into the quantum state of qubits using parameterized gates (e.g., rotation gates $R_x$, $R_y$, $R_z$) or amplitude embedding techniques. This transforms classical information into a manipulable quantum state  $\ket{\psi_{in}}$. Here, we first construct a classical ANN to reduce the dimensionality of the input features and then encode the reduced features into quantum states (See Figure \ref{fig:a1}). The design of subsequent VQC can be found in the following steps.

(i) A sequence of parameterized quantum gates (e.g., $U(\theta)=\prod_i R_{\alpha_i}(\theta_i)\cdot$ CNOT) is applied to entangle qubits and rotate their states. The variational quantum circuit can be seen in Figure \ref{fig:a1}, which consists of the basic structure of QNN. This step typically involves:

Entangling Layers: Controlled gates (e.g., CNOT, CZ) to generate quantum correlations.

Parametric Rotations: Single-qubit gates (e.g., $R_x(\theta)$, $R_y(\theta)$, $R_z(\theta)$ or their arbitrary combinations) with trainable parameters $\theta$.

These operations form a variational ansatz $Q(\theta)$. The circuit entangles qubits through gates like CNOT, interspersed with parameterized single-qubit rotations. This part is called a QNN layer. Several layers can be repeatedly stacked in a circuit to enhance expressivity.

(ii) Measurement and Decoding: the processed qubit states $\ket{\psi_{out}}=Q(\theta)\cdot\ket{\psi_{in}}$ are measured in a computational basis (e.g., Pauli-Z operators). Expectation values $\langle Z_i \rangle$ are decoded into classical outputs (e.g., probabilities, scalar values) for downstream tasks. Like the encoding step, output features may be high-dimensional, thus, we also construct a classical ANN to lift the dimensionality of the expectation values of VQC output (See Figure \ref{fig:a1}).

Diffident to classical ANN, the gradient of a variational quantum circuit (VQC) is calculated using the parameter-shift rule, a fundamental technique in quantum machine learning. For a single parameter quantum gate $U(\theta)=e^{-i\theta G/2}$, where $G$ is a Hermitian generator satisfying $G^2=I$, the gradient of an expectation value (i.e., loss function for VQC) $f(\theta)=\langle 0|U^\dagger(\theta)HU(\theta)|0\rangle$ is given by:

\begin{equation}
\frac{\partial f(\theta)}{\partial\theta} = \frac{\partial}{\partial\theta}\langle 0|U^\dagger(\theta)HU(\theta)|0\rangle = \frac{1}{2}\left[f\left(\theta+\frac{\pi}{2}\right) - f\left(\theta-\frac{\pi}{2}\right)\right]
\end{equation}
Where $H$ is the observable operator. For a VQC with parameter vector $\vec{\theta}=(\theta_1, \theta_2, \ldots, \theta_n)$, the gradient is computed component-wise:

\begin{equation}
\nabla f(\vec{\theta}) = \left(\frac{\partial f}{\partial\theta_1}, \frac{\partial f}{\partial\theta_2}, \ldots, \frac{\partial f}{\partial\theta_n}\right)
\end{equation}

where each component is given by:

\begin{equation}
\frac{\partial f(\vec{\theta})}{\partial\theta_i} = \frac{1}{2}\left[f\left(\theta_1, \ldots, \theta_i + \frac{\pi}{2}, \ldots, \theta_n\right) - f\left(\theta_1, \ldots, \theta_i - \frac{\pi}{2}, \ldots, \theta_n\right)\right]
\end{equation}
The key insight is that the gradient can be equivalently expressed as a difference of shifted parameter evaluations. This avoids complex analytical differentiation, and enables gradient estimation simply by shifting parameters in the original circuit, thereby offering a practical scheme for implementing gradient computation on real quantum hardware.

\subsection{AFC-oriented Quantum Reinforcement Learning}

\begin{figure}[H]
\centering
\includegraphics[width=1.0\textwidth]{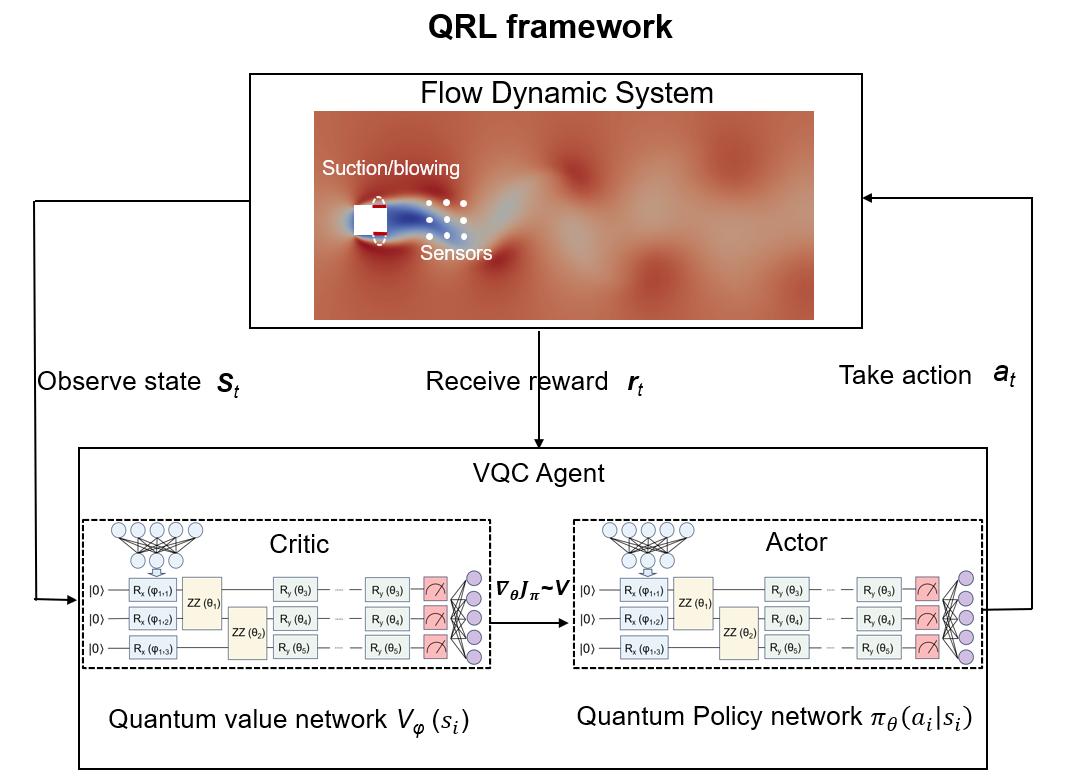}
\caption{QRL-based active flow control framework}
\label{fig:a2}
\end{figure}

In quantum reinforcement learning, VQC substitutes the policy training DNN of existing DRL. At each episode, the agent with given state information determines its action from policy-VQC and parameters are updated with a classical CPU algorithm like Adam Optimizer. This project materializes QRL, similarly, by using a VQC depicted in Figure \ref{fig:a2}. State encoding part (reduced-dimensional input features) of the circuit includes $R_x$ gates parameterized by normalized state input $s$, having their values between $-\pi$ and $\pi$. Variational part in the center consists of entangling CX Gates and $R_x$, $R_y$, $R_z$ gates parameterized with free parameter $\theta$. After that, measured output of the circuit yields the quantum state probabilities, and they are decoded into the action probabilities space via ANN. Then, the obtained $\pi_\theta$ is evaluated and updated in a classical computer.

In the beginning of an episode, quantum-classical hybrid agent receives state information from the environment (e.g., measured velocity of a flow dynamic system) and determines its action (flow control action such as blowing/suction) by $\pi_\theta$ made from the hybrid neural network. Then the policy of the agent is evaluated and updated by PPO algorithm, which is introduced before and described in Algorithm 1. The PPO algorithms are effectively the same as in the classical form. The replay buffer functions in the same way as in traditional approaches, keeping track of the $\langle s, a, R_t, s' \rangle$ tuples. For the optimization step, a classical optimizer (e.g., Adam, BFGS) computes gradients of a reward function $R_t(\theta)$ with respect to circuit parameters $\theta$. Parameters are updated iteratively ($\theta_{t+1} \leftarrow \theta_t - \eta \nabla L$) until convergence. This method enables the hybrid neural network ANN-VQCs to approximate complex functions, analogous to classical neural networks.

For the algorithm, we do not have to fundamentally or drastically change an algorithm in order to apply the power of VQC to it. The algorithm is presented in Algorithm 1.

\begin{algorithm}[H]
\caption{Quantum reinforcement learning with PPO}
\SetAlgoLined
\KwIn{Environment $E$, replay memory capacity $N$, total episodes $K$, discount factor $\gamma$, GAE parameter $\lambda$, PPO clipping parameter $\varepsilon$}
\KwOut{Trained quantum policy $\pi(\theta)$}
Initialize replay memory $D$ to capacity $N$\;
Initialize action-value quantum function $V(\phi)$ and quantum policy function $\pi(\theta)$\;
\For{episode = 1, 2, \ldots, K}{
  Initialize state $s_1$ and encode into quantum state through ANN (see Figure \ref{fig:a1})\;
  Collect set of trajectories $D_k$ by running quantum policy $\pi_k = \pi(\theta)$ in the environment\;
  Compute rewards-to-go $\hat{R}_t = \sum_{t'=t}^{T} \gamma^{t'-t} r_{t'}$\;
  Compute advantage estimates $A_{k}^{\pi_{\theta_k}} = \delta_k + (\gamma\lambda)\delta_{k+1} + \cdots + (\gamma\lambda)^{J-k+1}\delta_{J-1}$ based on the current value function $V_k$, where $\delta_k = r_k + \gamma V(s_{k+1}) - V(s_k)$\;
  Update the quantum policy by maximizing the PPO-Clip objective:
  \[
  \theta_{k+1} = \arg\max_{\theta} \frac{1}{|D_k|T} \sum_{\tau \in D_k} \sum_{t=0}^{T} \min\left(
  \frac{\pi_\theta(a_t \mid s_t)}{\pi_{\theta_k}(a_t \mid s_t)} A^{\pi_{\theta_k}}(a_t, s_t),\ 
  g(\varepsilon, A^{\pi_{\theta_k}}(a_t, s_t))
  \right)
  \]
  where $g(\varepsilon, A) = 
  \begin{cases}
  (1+\varepsilon)A & \text{if } A \geq 0 \\
  (1-\varepsilon)A & \text{if } A < 0
  \end{cases}$\;
  \Indp Typically, via stochastic gradient ascent with Adam\;
  \Indm
  Compute value function by regression on mean-squared error:
  \[
  \phi_{k+1} = \arg\min_{\phi} \frac{1}{|D_k|T} \sum_{\tau \in D_k} \sum_{t=0}^{T} \left( V_\phi(s_t) - \hat{R}_t \right)^2
  \]
  \Indp Typically, via same gradient descent algorithm\;
  \Indm
}
\Return Trained quantum policy $\pi(\theta)$\;
\end{algorithm}

\begin{itemize}
    \item $D$: Replay memory storing transitions $(s_t, a_t, r_t, s_{t+1})$
    \item $V(\phi)$: Quantum action-value function with parameters $\phi$
    \item $\pi(\theta)$: Quantum policy function with parameters $\theta$
    \item $\gamma$: Discount factor for future rewards
    \item $\lambda$: Parameter for Generalized Advantage Estimation
    \item $\varepsilon$: Clipping parameter for PPO
\end{itemize}

\subsection{Problems and numerical simulation setup}
\begin{figure}[H]
\centering
\includegraphics[width=0.8\textwidth]{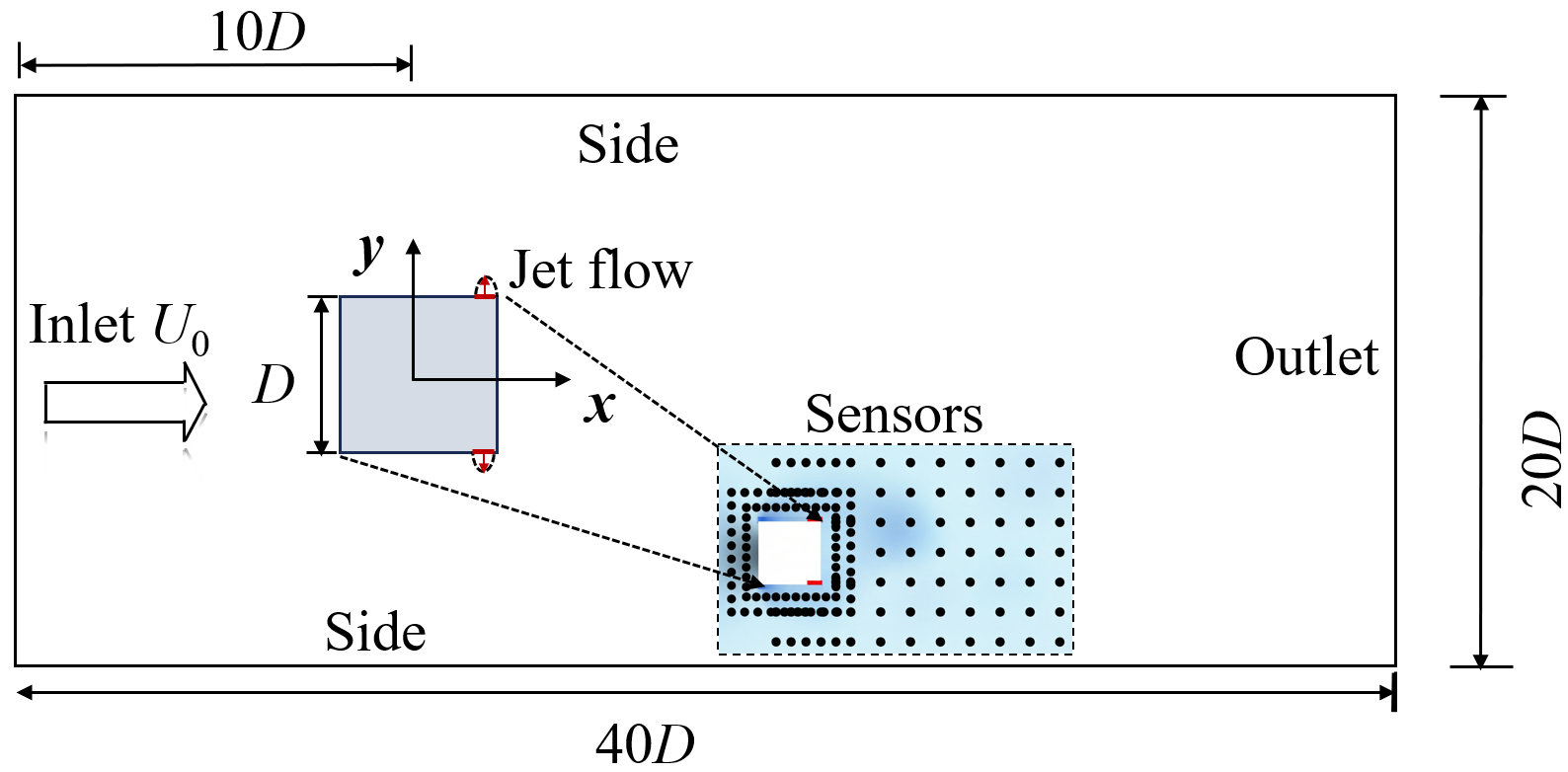}
\caption{The square cylinder and the computational domain}
\label{fig:a3}
\end{figure}

We demonstrate the QRL drag reduction framework on the flow past a 2-D square cylinder at laminar regimes characterised by 2-D vortex shedding. The case was performed at Reynolds numbers of $Re$ = 100 immersed in a computational domain. The diameter of the cylinder is $D$. Figure \ref{fig:a3} illustrates the computational domain and coordinate system of the numerical model. The origin of the Cartesian coordinate system is established at the center of the cylinder. The cylinder is immersed in a rectangular domain of dimensions 40$D$ (along the x-axis) and 20$D$ (along the y-axis). The distance from the inlet to the center of the square cylinder is 10 $D$, thus minimizing the effect of the outflow condition on vortex shedding. The Navier–Stokes equations can be written as

\begin{equation}
    \frac{\partial \mathbf{u}}{\partial t} + (\mathbf{u} \cdot \nabla) \mathbf{u} = -\frac{1}{\rho} \nabla p + \nu \nabla^2 \mathbf{u}
\end{equation}

\begin{equation}
    \nabla \cdot \mathbf{u} = 0
\end{equation}

\noindent where $p$ is the pressure and $u$ is the velocity vector.

The equations were solved using the finite volume element open-source code OpenFOAM. The non-dimensionalised inflow profile $U_0$ is uniform. On both sides of the domain, we apply the symmetrical boundary condition. For the square cylinder, a no-slip condition was employed on their surfaces. For pressure, zero normal gradient condition was employed at the boundary conditions of the inlet, two sides, and cylinders. For the outlet boundary, the flow velocity had a Neumann boundary condition with a zero normal gradient, and the pressure had a homogeneous Dirichlet boundary condition. We implemented a three-step time-splitting method with a velocity correction technique to attain second-order time accuracy. Structural meshes with more than 23000 quadrilateral elements are adopted for the CFD simulation, and they are refined around the square cylinder. A non-dimensional time step $dt$ = 0.0005 is adopted. The corresponding CFL number is less than unity, which meets typical accuracy requirements. We implemented implicit second-order scheme for time advance term.

The time-dependent lift $F_l$ and drag $F_d$ are obtained by integrating the instantaneous pressure and shear stress along the cylinder surface. And the lift coefficient and drag coefficient are defined as $C_L = F_l /0.5DU_0^2$ and $C_D = F_d /0.5DU_0^2$, respectively.

In this case, the reward function is set as:

\begin{equation}
    r = (C_D)_{T0} - C_D-r_0C_L'
\end{equation}

\noindent where $T_0$ is initial time. $r_0$ = 0.1, $C_D$ and $C'_L$ are the mean and standard deviation of the cylinder, respectively. $\omega$ is the rotational rate. The mean value of drag ($C_D$) and its fluctuation is always higher in magnitude than lift ($C'_L$) hence, the primary objective is to reduce $C_D$. Therefore, the weight for the drag is larger than the weight for lift. The total rewards at the end of the trajectory are computed as the discounted sum of rewards collected from every control time step. OpenFOAM acts as environment for fluid dynamics control. Our experiment is conducted with the software packages, PyTorch for speed and convenience of tensor operation and PennyLane for quantum circuit simulation. We used classical parameter optimizer as Adam optimizer with learning rate 0.001 for quantum policy and 0.00001 for classical critic. Other hyperparameter settings are $\gamma = 0.98$, $\lambda = 0.95$, $\epsilon = 0.01$.

\begin{figure}[H]
\centering
\includegraphics[width=0.8\textwidth]{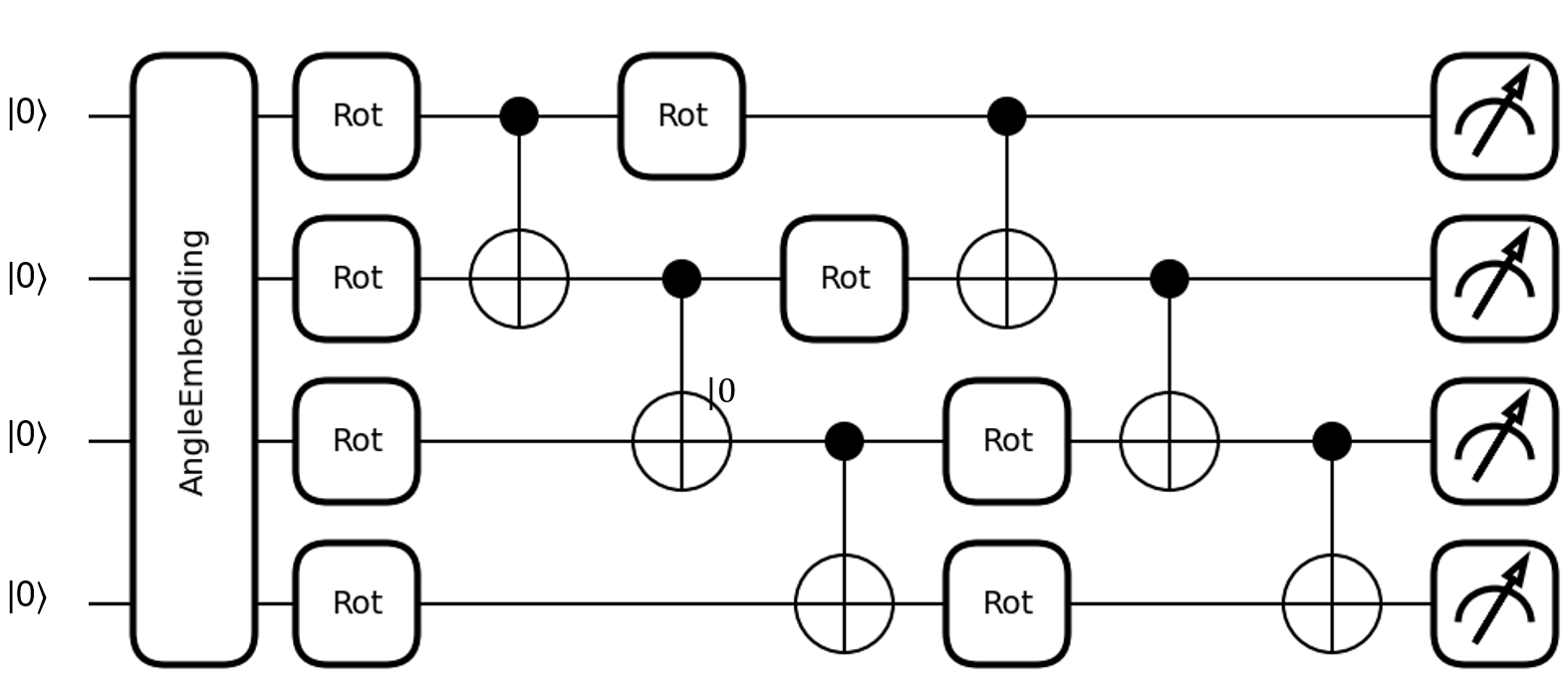}
\caption{VQC of actor and critic network in the CartPole problem}
\label{fig:a4}
\end{figure}

\section{Results and Discussion}

\begin{figure}[H]
\centering
\includegraphics[width=1.0\textwidth]{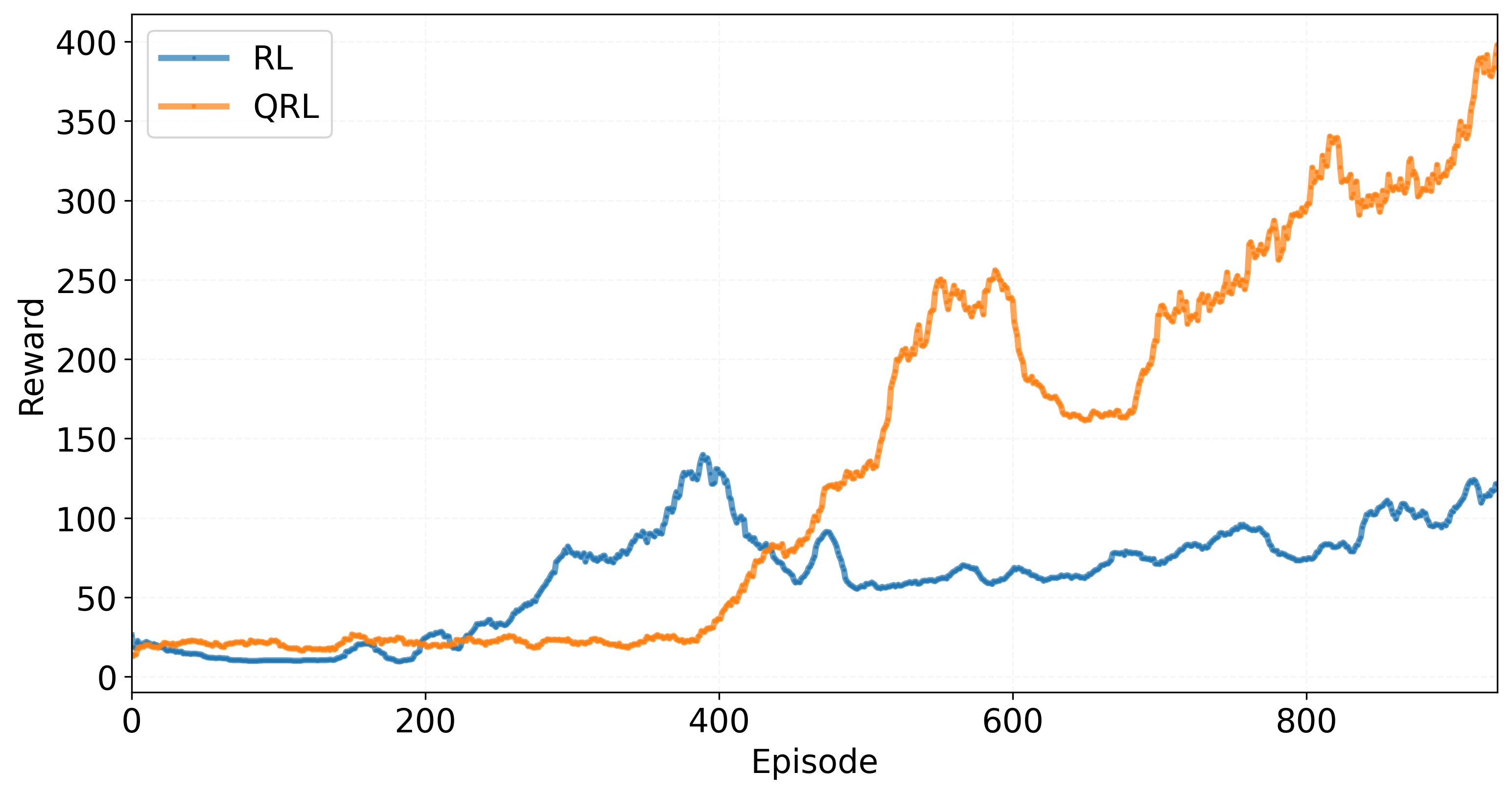}
\caption{Comparison of the performance between RL and QRL in the CartPole problem}
\label{fig:a5}
\end{figure}

\subsection{A toy model test}

The CartPole environment, a reinforcement learning benchmark provided by OpenAI, simulates a control task where an agent must balance a pole attached to a moving cart. The agent learns to apply forces to move the cart left or right, aiming to prevent the pole from falling for as long as possible. A positive reward of +1 is granted for each time step the pole remains upright, and the episode terminates when the pole exceeds a specific angle or the cart moves too far from the center. The state observed by the agent consists of four continuous variables: the cart's position, the cart's velocity, the pole's angle, and the pole's angular velocity. Based on this state information, the agent selects an action to accelerate the cart accordingly. In this context, the VQC shown in Figure \ref{fig:a4}, which utilizes 4 qubits, is appropriate for policy formulation in this environment. Each of the four state variables is normalized to values between $-\pi$ and $\pi$ before being input into the circuit through angle embedding. The two measurement outcomes from the circuit are then decoded into probability values for the two possible actions using a softmax function. This interaction cycle continues as the agent optimizes its parameters through a RL algorithm to identify optimal actions for given states. 

For parameter optimization, the quantum policy was trained using the Adam optimizer with a learning rate of 0.001, while the classical critic used Adam with a learning rate of 0.00001. Additional hyperparameters were set as follows: discount factor $\gamma$ = 0.98, trace decay $\lambda$ = 0.95, and clipping threshold $\epsilon$ = 0.01. As a baseline, we also evaluated a randomized version of the model, which applies untuned random parameters at each time step without any optimization.

\begin{table}[htbp]
\centering
\caption{Parameter Comparison between QRL and Classical RL Models}
\begin{tabular}{lccc}
\toprule
Network Type & Policy Network & Value Network & Total \\
            & Parameters     & Parameters    & Parameters \\
\midrule
QRL & 398  & 365  & 763 \\
RL  & 4,610 & 4,545 & 9,155 \\
\bottomrule
\end{tabular}
\label{tab:param_comp}
\end{table}

Figure \ref{fig:a5} illustrates the performance of the proposed QRL and classical RL model. The comparative analysis demonstrates substantial advantages in both computational efficiency and learning performance for the quantum-enhanced approach. As evidenced in the parameter comparison, the QRL achieves remarkable model compression, with a 91.7\% reduction in total parameters (763 vs. 9,155 parameters) while maintaining competitive or superior performance. This significant parameter efficiency translates to reduced memory footprint and computational overhead, making the hybrid architecture particularly advantageous for resource-constrained applications. The performance trajectory, as illustrated in the reward evolution plot, further validates the effectiveness of the QRL, which exhibits accelerated convergence and achieves higher cumulative rewards (often exceeding 450) compared to the classical counterpart that struggles to consistently surpass the 200-reward threshold.

The learning dynamics reveal fundamentally different behavioral patterns between the two approaches. The QRL demonstrates not only faster learning but also greater stability during training stages. In contrast, the RL exhibits slower convergence throughout the training period. This performance suggests that the quantum circuit component can enhance the model's representational capacity and optimization efficiency, potentially through quantum superposition and entanglement mechanisms that facilitate more effective exploration of the policy space.
 
Although the case studied is relatively simplified and may lack generality, our findings still suggest the promising potential of QRL. Specifically, a VQC with notably fewer parameters can exhibit expressive power surpassing that of a classical neural network with orders of magnitude more parameters.

\subsection{QRL-based control on flow past square cylinder}
\begin{figure}[H]
\centering
\includegraphics[width=1.0\textwidth]{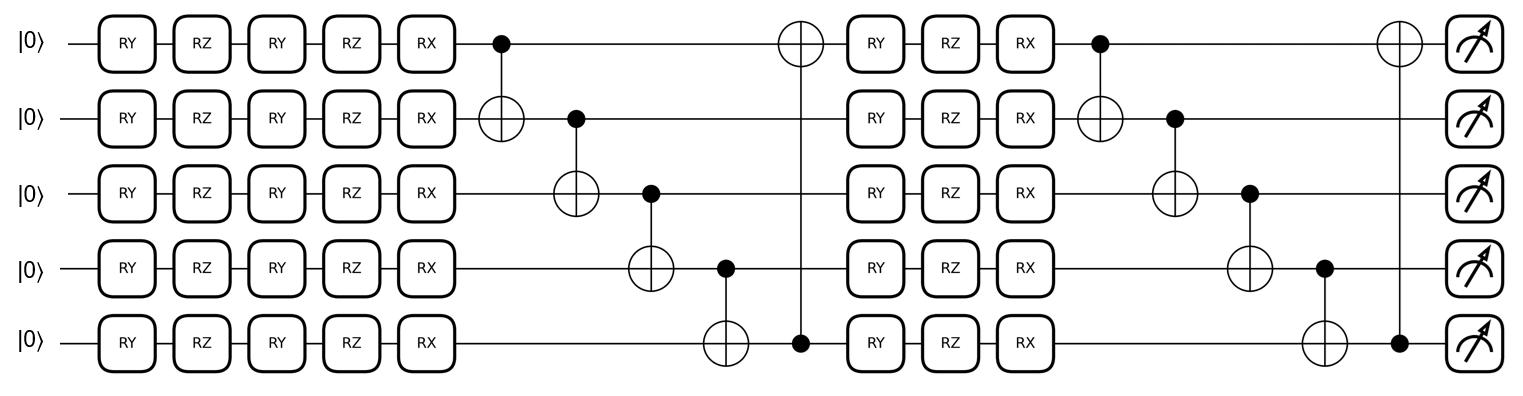}
\caption{VQC of actor and critic network in the QRL-based flow control}
\label{fig:a6}
\end{figure}

\begin{figure}[H]
\centering
\includegraphics[width=0.8\textwidth]{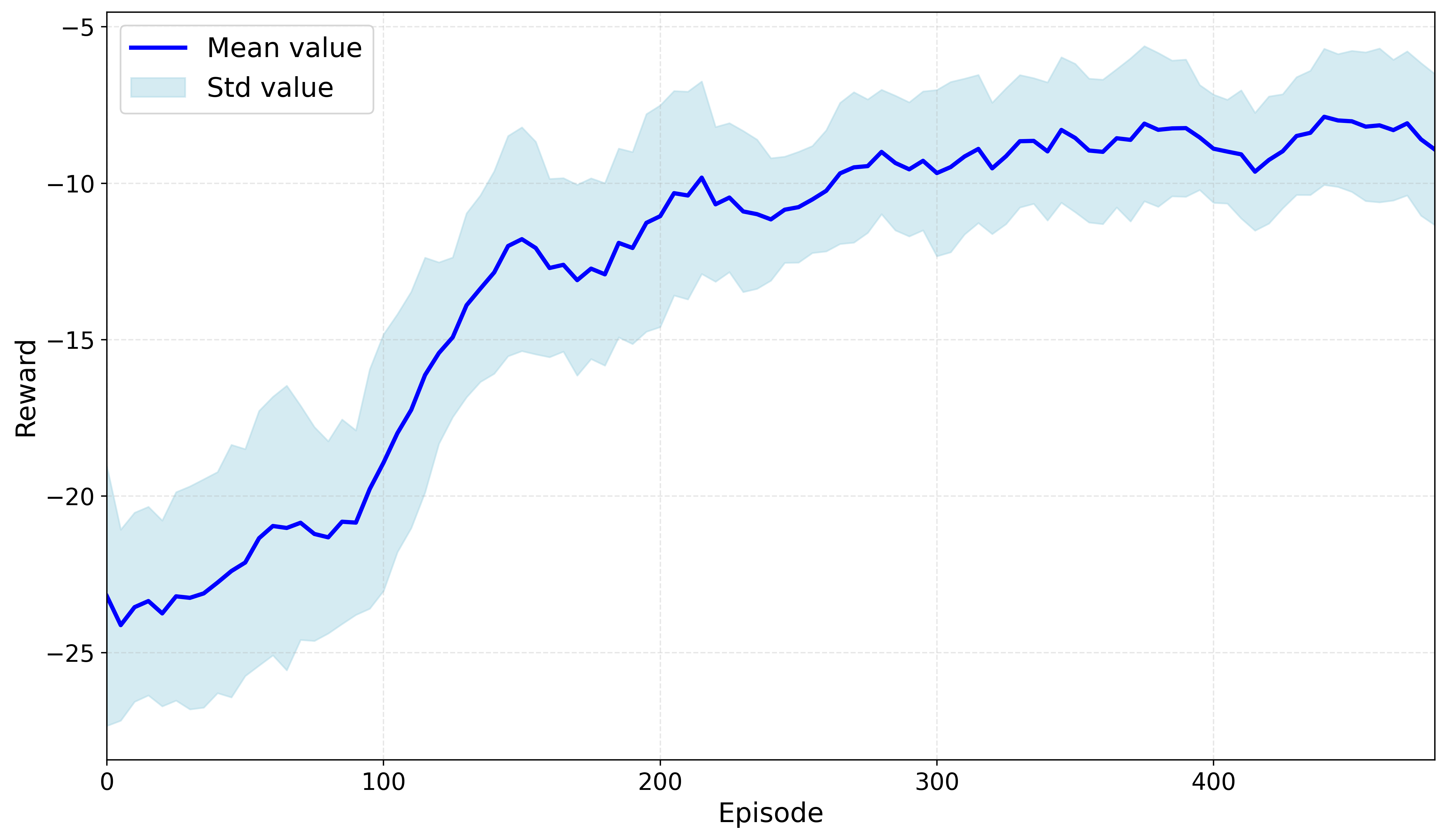}
\caption{Performance of the flow control via QRL}
\label{fig:a7}
\end{figure}

In flow control case, both the actor and critic networks employ the same VQC architecture. As shown in Figure \ref{fig:a6}, a single quantum layer of this network consists of 5 qubits. The network structure includes quantum rotations and entanglement operations. Both the actor and critic networks comprise 2 layers of such variational quantum circuits. Figure \ref{fig:a7} presents the learning curve, depicting the evolution of the reward as a function of training episodes for a QRL agent applied to a flow control task. The plot illustrates the characteristic learning progression of the agent, where the mean reward (solid blue line) demonstrates a clear, monotonic increase from its initial value towards a higher asymptotic level. This positive trajectory indicates the successful convergence of the QRL policy, as the agent systematically improves its control strategy to maximize the defined reward signal over time. Concurrently, the shaded region representing the standard deviation (Std value) around the mean shows notable fluctuation, particularly in the early to mid-phase of training (episodes ~50-300). This pattern reflects the inherent exploration-exploitation trade-off and the stochastic learning dynamics; the initially higher variance suggests active exploration of the state-action space, which gradually reduces as the policy stabilizes towards a more deterministic, high-reward solution. The eventual convergence of both the mean and a bounded standard deviation to a steady level signifies that the QRL algorithm has learned a robust and effective control policy for the given task, highlighting its capability in handling the complexities of the flow control environment. 

\begin{figure}[H]
\centering
\includegraphics[width=0.8\textwidth]{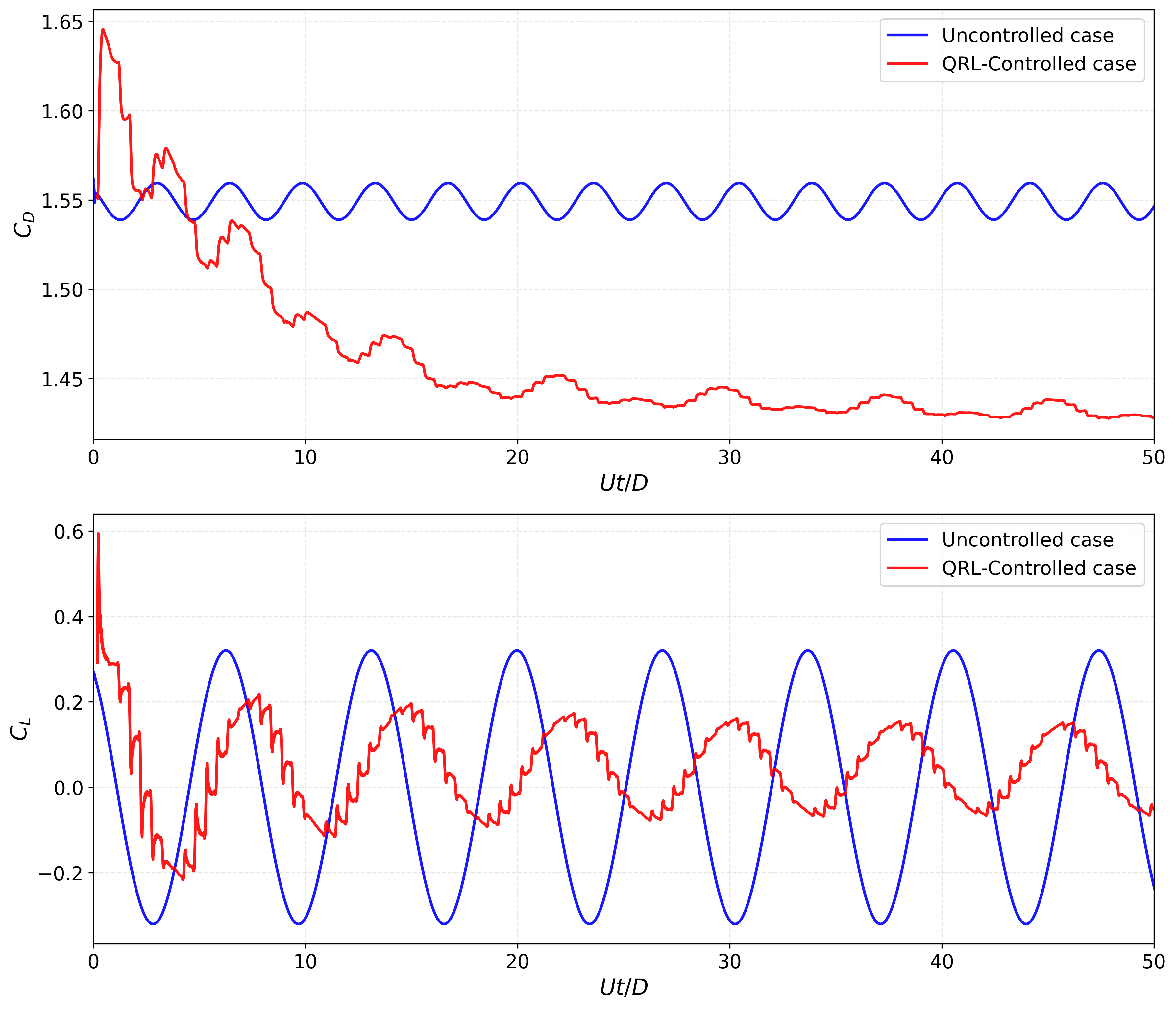}
\caption{(a) Drag and (b) lift coefficients under QRL-controlled and uncontrolled case}
\label{fig:a8}
\end{figure}

Figure \ref{fig:a8} presents the time histories of the drag and the lift coefficient for flow past a square cylinder, comparing the uncontrolled case with the closed-loop controlled by the QRL agent. In Figure \ref{fig:a8} (a), the $C_D$ of uncontrolled case remains stable periodic oscillation, and the man value is approximately 1.55, consistent with the results of Wang et al. (2022). In contrast, the QRL-controlled $C_D$ exhibits a lower mean value of 1.41, with weakly fluctuating oscillation, which demonstrates the controller's primary objective of effective drag reduction. This confirms that the control strategy not only mitigates lift oscillations but also achieves a substantial net reduction in the mean drag, highlighting a comprehensive performance improvement. Figure \ref{fig:a8} (b) highlights the controller's effect on the unsteady lift force. The uncontrolled $C_L$ shows large-amplitude, regular periodic oscillations, characteristic of the von Kármán vortex street. The QRL control successfully suppresses these lift fluctuations, drastically reducing the oscillation amplitude. The amplitude of $C_L$ is reduced from about 0.3 to 0.12, and the waveform becomes more irregular, indicating a direct modification of the vortex shedding dynamics. In summary, the QRL-based controller successfully achieves its dual objectives: it reduces the mean drag force while simultaneously attenuating the unsteady lift fluctuations, confirming its capability for effective flow control. 

\begin{figure}[H]
\centering
\includegraphics[width=0.6\textwidth]{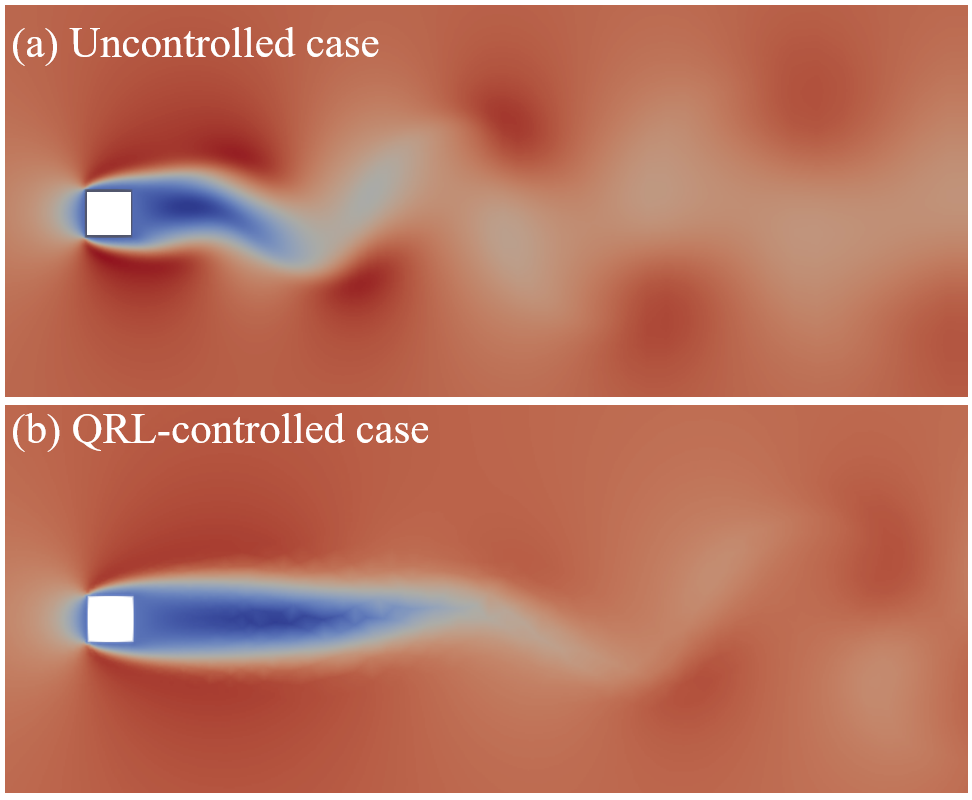}
\caption{Flow field for (a) baselines case and (b) QRL control)}
\label{fig:a9}
\end{figure}

Figure \ref{fig:a9} presents a comparison of the instantaneous streamwise velocity for the uncontrolled baseline case Figure \ref{fig:a9} (a) and the flow under QRL control Figure \ref{fig:a9} (b). In the baseline case, the classic von Kármán vortex street is clearly visible downstream of the cylinder. This flow is characterized by the periodic shedding of large-scale, coherent vortices of alternating sign, which are the primary source of significant unsteady drag and lift forces on the cylinder. The sharp, well-defined vortical structures and the wide wake region indicate a highly energetic and unstable flow regime. QRL-based flow control shows a profound modification of the near-wake dynamics, as shown in Figure \ref{fig:a9} (b). The most striking effect is the substantial attenuation of the large-scale vortex shedding. The coherent vortical structures observed in the baseline are effectively broken down and suppressed. The wake appears noticeably narrower and more streamlined, with the vorticity field displaying smaller spatial scales and a significantly reduced intensity. This visual transformation indicates a successful stabilization of the wake flow, leading to a less energetic and more ordered flow pattern immediately behind the square cylinder. This dramatic alteration in the vortex dynamics visually corroborates the quantitative performance improvements discussed earlier. The suppression of the powerful periodic vortex shedding directly explains the reduction in mean drag and the effective damping of the large-amplitude lift fluctuations, as the primary source of these forces is mitigated by the QRL control.

\section{Conclusions}
This study presents a novel QRL framework for active flow control applications. The proposed hybrid architecture, which integrates VQC with a classical PPO algorithm, is first evaluated on the benchmark CartPole problem. Comparative analysis reveals that the QRL agent achieves a remarkable reduction in the number of trainable parameters compared to its classical RL counterpart, while attaining superior performance. This signifies a substantial potential improvement in parameter efficiency and suggests a stronger representation capacity of the quantum policy network, making it particularly suitable for complex fluid dynamic applications.

Subsequently, the QRL framework is applied to control the flow past a square cylinder at a Reynolds number of 100, employing real-time blowing/suction actuation on the cylinder’s surface. The controller successfully derives an effective policy that significantly alters the wake dynamics. Results confirm that the QRL-based control effectively suppresses large-scale vortex shedding, leading to a narrower and more stable wake structure. Consequently, it achieves a reduction in the mean drag coefficient and a substantial attenuation of the unsteady lift fluctuations. These results present the capability of the QRL framework in managing complex, nonlinear fluid systems.

In summary, this work establishes the potential of quantum-enhanced reinforcement learning as a powerful tool for tackling high-dimensional, nonlinear control challenges in fluid dynamics. The combination of parameter efficiency and robust control performance positions the QRL framework as a promising paradigm for future applications in aerospace engineering, turbomachinery, and other fields involving complex fluid-structure interactions. Future research will focus on extending this approach to higher Reynolds number turbulent flows and experimental validations, further exploring the scalability and practical robustness of quantum-classical hybrid algorithms in real-world settings.

\section{References}
\begin{enumerate}[label={}]

\item Bergmann, M., Cordier, L., \& Brancher, J.-P. (2005). Optimal rotary control of the cylinder wake using proper orthogonal decomposition reduced-order model. Physics of Fluids, 17(9), 097101.
\item Brackston, R. D., Garcia de la Cruz, J. M., Wynn, A., Rigas, G., \& Morrison, J. F. (2016). Stochastic modelling and feedback control of bistability in a turbulent bluff body wake. Journal of Fluid Mechanics, 802, 726–749.
\item Brunton, S. L., \& Noack, B. R. (2015). Closed-loop turbulence control: Progress and challenges. Applied Mechanics Reviews, 67(5), 050801.
\item Brunton, S. L., Noack, B. R., \& Koumoutsakos, P. (2020). Machine learning for fluid mechanics. Annual Review of Fluid Mechanics, 52, 477–508.
\item Caro, M. C., Huang, H.-Y., Cerezo, M., Sharma, K., Sornborger, A., Cincio, L.,\& Coles, P. J. (2022). Generalization in quantum machine learning from few training data. Nature Communications, 13(1), 4919.
\item Chen, Z., \& Aubry, N. (2005). Active control of cylinder wake. Journal of Fluid Mechanics, 529, 27–44.
\item Dong, D., Chen, C., Li, H., \& Tarn, T. J. (2008). Quantum reinforcement learning. IEEE Transactions on Systems, Man, and Cybernetics - Part B: Cybernetics, 38(5), 1207–1220.
\item Dong, D., Chen, C., Chu, J., \& Tarn, T.-J. (2012). Robust quantum-inspired reinforcement learning for robot navigation. IEEE/ASME Transactions on Mechatronics, 17(1), 86–97.
\item Dunjko, V., Taylor, J. M., \& Briegel, H. J. (2016). Quantum-enhanced machine learning. Physical Review Letters, 117(13), 130501.
\item Duraisamy, K., Iaccarino, G., \& Xiao, H. (2019). Turbulence modeling in the age of data. Annual Review of Fluid Mechanics, 51, 357–377.
\item Flinois, T. L. B., \& Colonius, T. (2015). Optimal control of circular cylinder wakes using long control horizons. Physics of Fluids, 27(8), 087105.
\item Garnier, P., Viquerat, J., Rabault, J., Larcher, A., Kuhnle, A., \& Hachem, E. (2021). A review on deep reinforcement learning for fluid mechanics. Computers \& Fluids, 225, 104973.
\item Ghiglieri, J., \& Ulbrich, S. (2014). Optimal flow control based on POD and MPC and an application to the cancellation of Tollmien–Schlichting waves. Optimization Methods and Software, 29(5), 1042–1074.
\item Krizhevsky, A., Sutskever, I., \& Hinton, G. E. (2017). ImageNet classification with deep convolutional neural networks. Communications of the ACM, 60(6), 84–90.
\item Leclercq, C., Demourant, F., Poussot-Vassal, C., \& Sipp, D. (2019). Linear iterative method for closed-loop control of quasiperiodic flows. Journal of Fluid Mechanics, 868, 26–65.
\item LeCun, Y., Bengio, Y., \& Hinton, G. (2015). Deep learning. Nature, 521(7553), 436–444.
\item Levine, S., Pastor, P., Krizhevsky, A., Ibarz, J., \& Quillen, D. (2018). Learning hand-eye coordination for robotic grasping with deep learning and large-scale data collection. The International Journal of Robotics Research, 37(4-5), 421–436.
\item Li, J. A., Dong, D., Wei, Z., Liu, Y., Pan, Y., Nori, F., \& Zhang, X. (2020). Quantum reinforcement learning during human decision-making. Nature Human Behaviour, 4(3), 294–307.
\item Li, Y., Noack, B. R., \& Cordier, L. (2006). Genetic algorithm-based active control of a circular cylinder wake. Journal of Turbulence, 7, N52.
\item Mnih, V., Kavukcuoglu, K., Silver, D., Rusu, A. A., Veness, J., Bellemare, M. G., ... Hassabis, D. (2015). Human-level control through deep reinforcement learning. Nature, 518(7540), 529–533.
\item Razvarz, S., Vargas-Jarillo, C., Jafari, R., \& Gegov, A. (2019). Flow control of fluid in pipelines using PID controller. IEEE Access, 7, 25673–25680.
\item Schanz, D., Gesemann, S., \& Schröder, A. (2016). Shake-The-Box: Lagrangian particle tracking at high particle image densities. Experiments in Fluids, 57(5), 70.
\item Schoppa, W., \& Hussain, F. (1998). A large-scale control strategy for drag reduction in turbulent boundary layers. Physics of Fluids, 10(5), 1049–1051.
\item Tai, W., Yang, Y., \& Zaki, T. A. (2021). Data-enhanced turbulence modeling with neural networks. Physical Review Fluids, 6(11), 114604.
\item Viquerat, J., Meliga, P., Larcher, A., \& Hachem, E. (2022). A review on deep reinforcement learning for fluid mechanics: An update. Physics of Fluids, 34(11).
\item Wang, M., Li, Y., \& Noack, B. R. (2011). Genetic algorithm-based optimization of active flow control. Theoretical and Applied Mechanics Letters, 1(4), 041001.
\item Wang, Q., Yan, L., Hu, G., Li, C., Xiao, Y., Xiong, H., Rabault, J. \& Noack, B.R. (2022). DRLinFluids: An open-source Python platform of coupling deep reinforcement learning and OpenFOAM. Physics of Fluids, 34(8).
\item Wieneke, B. (2017). PIV uncertainty quantification from correlation statistics. Measurement Science and Technology, 28(7), 075301.
\item Zhang, H., Li, Y., \& Noack, B. R. (2013). Genetic algorithm-based optimal control of fluid flow. Journal of Fluid Mechanics, 729, 1–35.
\item Sutton, R. S., \& Barto, A. G. (2018). Reinforcement learning: An introduction(2nd ed.). MIT Press.
\end{enumerate}

\end{document}